\documentstyle[12pt]{article}

\def\be{\begin{equation}}
\def\ee{\end{equation}}
\def\a{\alpha}
\def\b{\beta}
\def\ra{\rangle}
\def\la{\langle}
\def\bt{{\tilde b}} 
\def\Nh{\hat{N}}
 
\def\ctg{\mbox{ctg}} 
\def\sh{\mbox{sh}}
\def\ch{\mbox{ch}}
\def\sin{\mbox{sin}}
\def\cos{\mbox{cos}}
\def\exp{\mbox{exp}} 
\def\g{\gamma}
\def\w{\omega}

\begin{document}
 
\begin{center}
{\bf Lattice local integrable regularization of the Sine-Gordon model.}
\end{center}
\vspace{0.2in}
\begin{center}

{\large A.A.Ovchinnikov}

\end{center}   

\begin{center}
{\it Institute for Nuclear Research, RAS, Moscow}
\end{center}   
 
\vspace{0.2in}

\begin{abstract}

We study the local lattice integrable regularization of the Sine-Gordon 
model written down in terms of the lattice Bose- operators. 
We show that the local spin Hamiltonian obtained from the six-vertex model 
with alternating inhomogeneities in fact leads to the Sine-Gordon in the 
low-energy limit. We show that the Bethe Ansatz results for this model 
lead to the correct general relations for different critical exponents 
of the coupling constant.

\end{abstract}

\vspace{0.2in}

{\bf 1. Introduction}

\vspace{0.2in}

It is interesting and important to study various integrable lattice 
regularizations of the integrable quantum field theory models in 
two dimensions. Among them the lattice regularizations connected 
with the vertex models associated with the trigonometric $S$-matrix 
are especially interesting. In particular for the Sine-Gordon (SG) 
model the so called Light Cone lattice approach was proposed \cite{DV}. 
The main shortcoming of this lattice Hamiltonian is its non-locality. 
The local version of this Hamiltonian was proposed in \cite{RS} and 
later its Fermionic version was studied in Ref.\cite{DS}.   
However to reduce the Lagrangian to the Lagrangian of the SG- model, 
the naive Bosonization of the lattice strongly-interacting Fermionic 
operators was used, so that even the parameter $\b$ in terms of the 
parameters of the six-vertex model was not calculated directly. 

In the present Letter we use directly the Bosonic version of the 
approaches \cite{RS},\cite{DS} which allows one to deal with the 
original (bosonic) six-vertex model with the alternating 
inhomogeneity parameters. We reduce the original lattice problem to 
the system of two weakly coupled XXZ- spin chains and perform the 
Bosonization rigorously, 
which allows us to obtain the Hamiltonian of the SG- model directly 
without any reference to the Massive Thirring Model. 
In particular we calculate the constant $\beta$ - directly from the 
well defined Bosonization procedure. 

We write down the lattice Hamiltonian and the Bethe Ansatz (BA) 
equations in Section 2. We perform the Bosonization of this 
Hamiltonian using the well known formulas for the spin operators in 
the XXZ- spin chain in Section 3. Finally in Section 4 we check the 
general relations for the critical exponents in the expansion of 
the physical quantities in the coupling constant.

\vspace{0.2in}

{\bf 2. Lattice Hamiltonian.}

\vspace{0.2in}

To write down the Hamiltonian let us first introduce the well known 
trigonometric $S$- matrix. In the simplified notations it has the form 
\[
S_{12}(t_1-t_2)=(\sh(t+i\eta),~ \sh(t),~ \sh(i\eta))_{12},~~~ t=t_1-t_2, 
\]
and obeys the Yang-Baxter equation $S_{12}S_{13}S_{23}=S_{23}S_{13}S_{12}$. 
The transfer matrix $Z(t)$ acting in the quantum space $(1,\ldots L)$ 
has the form:
\be
Z(t)=Z(t;\xi_1,\ldots\xi_L)=Tr_0
\left(S_{10}S_{20}\ldots S_{L0}\right), 
\label{Z}
\ee
where the inhomogeneity parameter $\xi_i$ corresponds to each site $i$. 
We choose the alternating values of the parameters $\xi_{2k+1}=0$, 
$\xi_{2k}=\theta$, $k\in Z$. The local Hamiltonian corresponding to the 
SG- model equals 
\be
H=H(0)+H(\theta)=\frac{\sh(i\eta)}{2}
\left(Z^{-1}(0)\dot Z(0)+Z^{-1}(\theta)\dot Z(\theta)\right), 
\label{H} 
\ee
where the dots stand for the derivatives. 
Substituting the transfer matrix (\ref{Z}) into the equation (\ref{H}) 
we obtain the following local Hamiltonian: 
\be 
H=\sum_{i}\left(S_{i+1,i+2}^{-1}P_{i,i+2}\dot S_{i,i+2}S_{i+1,i+2}+ 
\sh(i\eta)S_{i+1,i+2}^{-1}\dot S_{i+1,i+2}\right), 
\label{Ham}
\ee
where the periodic boundary conditions are implied and $P_{ij}$ is the 
permutation operator. The Hamiltonian (\ref{Ham}) is correct both for 
even and odd $i$ provided the corresponding inhomogeneity parameter 
is the spectral parameter for each site. 
We consider the Hamiltonian (\ref{Ham}) at large $\theta$ and expand 
it in powers of $e^{-\theta}$. At $e^{-\theta}=0$ we get two coupled 
XXZ- spin chains: 
\be
H_0=\frac{1}{2}b_1^{+}b_3e^{i2\eta(n_2-1/2)}+h.c.+\Delta n_1n_3+ 
\frac{1}{2}b_2^{+}b_4e^{-i2\eta(n_3-1/2)}+h.c.+\Delta n_2n_4+\ldots, 
\Delta=\cos(\eta), 
\label{H0}
\ee 
where the dots stand for the next terms of the odd and even spin chains 
and the hard-core bosons $b_i^{+}$, $b_i$, $n_i=b_i^{+}b_i$ are introduced 
to describe the state at the site $i$ in such a way that $n_i=1$ ($n_i=0$) 
corresponds to the spin-up (spin-down) state and $b_i^{+}$ ($b_i$) change 
the direction of spin. 
Now we can remove the interaction between two chains performing the 
transformation
\be
\bt_{1x}^{+}=b_{1x}^{+}e^{i2\eta\Delta N_2(x)}, ~~~
\bt_{2x}^{+}=b_{2x}^{+}e^{-i2\eta\Delta N_1(x)},
\label{trans}
\ee
where the notations $b_{1x}^{+}=b_{2x-1}^{+}$, $b_{2x}^{+}=b_{2x}^{+}$, 
$x=1,2,\ldots L/2$ are used and $\Delta N_1(x)=\sum_{i<x}(n_{1i}-1/2)$, 
$\Delta N_2(x)=\sum_{i<x}(n_{2i}-1/2)$. In terms of the new operators 
$\bt_{1x}^{+}$, $\bt_{2x}^{+}$ (\ref{trans}) the Hamiltonian (\ref{H0}) 
takes the form of two independent XXZ- spin chains and one can use the 
known results for the spin chain to study the Hamiltonian (\ref{Ham}). 
Note that this substitution leads to the kind of the twisted boundary 
conditions which enter the effective low-energy theory through the 
well defined quantum number $\Delta N_1+\Delta N_2=M-L/2$ which show 
which boundary conditions for the field $\phi(x)$ $\sim(\phi(L/2)-\phi(0))$ 
in the Lagrangian (\ref{SG}) below are actually implied (see Section 3). 
The Bethe Ansatz equations (see eq.(\ref{BA}) below) 
will take care about these twists automatically. 
Now we calculate the interaction of the two spin chains of order $e^{-\theta}$. 
The explicit form of the Hamiltonian (\ref{Ham}) is rather complicated. 
The task is simplified if one is extracting the terms which lead to the 
relevant interaction $\sim\cos(\beta\phi)$. For example, the terms which 
contain the factor $(n_i-1/2)\sim\partial_{x}\phi(x)|_{x=i}$, where 
$\phi(x)$- is some Bose field, lead to the irrelevant operators and 
can be omitted. Analogously the factors $n_i$ in front the operators 
$b_{i}^{+}$, $b_{i}$ can be substituted as $n_i\rightarrow 1/2$. 
The result of the calculations has the form: 
\be
\hat{V}=2(\sin(\eta))^{2}e^{-\theta}\left(\sum_{x}(b_{1x}^{+}b_{2x}+h.c.)+ 
\sum_{x}(b_{1(x+1)}^{+}b_{2x}+h.c.)\right). 
\label{V} 
\ee
From the point of view of application of the Bosonization procedure 
the two terms in eq.(\ref{V}) are very similar and the factor $2$ comes 
from the two different terms in eq.(\ref{Ham}). Below we will use (\ref{V}) 
to derive the SG model.

\vspace{0.2in}

{\bf 3. Bosonization.}

\vspace{0.2in}

We have seen that up to the order $\sim e^{-\theta}$ the Hamiltonian 
has the form of two coupled XXZ- spin chains. The low-energy 
effective theory for an XXZ- spin chain is well known: it is the 
Luttinger liquid (for example, see \cite{LP},\cite{H}) with the 
parameter $\xi=2(\pi-\eta)/\pi$, which after rescaling of $x$ and $t$ 
is equivalent to the free massless Bose field. For the two chains 
we have to such fields. Now we consider the interaction of two 
chains and seek for the operators which are {\it relevant} and 
neglect the operators which are irrelevant in the low-energy limit. 
To do it we perform the Bosonization of the Hamiltonian (\ref{Ham}). 
The analysis performed in the previous Section 
shows that the only relevant interaction term has the 
form (\ref{V}) which can be further simplified to 
\be
\hat{V}=hC\sum_{x}b_{1x}^{+}b_{2x}+h.c.,  
\label{int}
\ee
where now $C$ is some unimportant constant (see eq.(\ref{V})) and 
where the coupling constant $h=e^{-\theta}$. To express this interaction 
in term of the scalar Bose fields one can use the well known Bosonization 
formulas for the XXZ- spin chain. We have up to the constant: 
\be
\bt_{1x}^{+}\simeq (-1)^{x}e^{-i\pi\sqrt{\xi}(\Nh_1-\Nh_2)^{(1)}(x)}, ~~~~
\bt_{2x}\simeq (-1)^{x}e^{i\pi\sqrt{\xi}(\Nh_1-\Nh_2)^{(2)}(x)}, 
\label{oper}
\ee
where the operators $\Nh_{1,2}^{(i)}(x)$ for the two chains $i=1,2$ are 
expressed through the initial Fermi-operators of the Luttinger model 
$a_{1,2}(k)$ as 
\[
\Nh_{1}-\Nh_{2}=(1/\sqrt{\xi})(N_1-N_2), ~~~~
\Nh_{1}+\Nh_{2}=\sqrt{\xi}(N_1+N_2), 
\]
where $\xi$- is the standard Luttinger liquid parameteter and  
\[
N_{1,2}(x)=\frac{i}{L^{\prime}}\sum_{p\neq 0}\frac{\rho_{1,2}(p)}{p}e^{-ipx}, 
~~~~\rho_{1,2}(p)=\sum_{k}a_{1,2}^{+}(k+p)a_{1,2}(k).  
\]
where $L^{\prime}=L/2$.
The standard Bose fields $\phi_1(x)$, $\phi_2(x)$ are connected with the 
fields $\Nh_{1,2}^{(i)}(x)$ in the following way: 
\[
(\Nh_{1}^{(i)}+\Nh_{2}^{(i)})(x)=(1/\sqrt{\pi})\phi_i(x), ~~~~
(\Nh_{1}^{(i)}-\Nh_{2}^{(i)})(x)=(1/\sqrt{\pi}){\tilde \phi}_i(x), ~~~i=1,2, 
\]
where the dual fields ${\tilde \phi}_i(x)$ are defined according to the 
equations 
\[
{\tilde \phi}_i(x)=\int^{x}dy\pi_i(y), ~~~\pi_i(x)=\dot \phi_i(x), ~~~i=1,2, 
\]
where $\pi_i(x)$- are the conjugated momenta. 
Now from the equations (\ref{trans}) one can see that the operator 
$b^{+}_{1x}$ take the following form: 
\be
b_{1x}^{+}\simeq \exp\left(-i\pi\sqrt{\xi}(\Nh_1-\Nh_2)^{(1)}(x)+
i(2\pi-2\eta)\frac{1}{\sqrt{\xi}}(\Nh_1+\Nh_2)^{(2)}(x)\right)=
e^{i\sqrt{\pi}\sqrt{\xi}({-\tilde \phi}_1(x)+\phi_2(x))}, 
\label{b1}
\ee
where the value $\xi=2(\pi-\eta)/\pi$ was substituted. Analogously for 
the operator $b_{2x}$ we obtain the expression 
\be
b_{2x}\simeq 
\exp\left(i\pi\sqrt{\xi}(\Nh_1-\Nh_2)^{(2)}(x)+
i(2\pi-2\eta)\frac{1}{\sqrt{\xi}}(\Nh_1+\Nh_2)^{(1)}(x)\right)=
e^{i\sqrt{\pi}\sqrt{\xi}({\tilde  \phi}_2(x)+\phi_1(x))}. 
\label{b2}
\ee
Note that in the process of the derivation of (\ref{b1}), (\ref{b2}) 
we have inserted the additional factors equal to unity of the form 
$\prod_{i<x}e^{i2\pi n_i}=(-1)^{x}\prod_{i<x}e^{i2\pi(n_i-1/2)}$ 
which cancels the factors $(-1)^{x}$ in the equations (\ref{oper}). 
Combining the equations (\ref{b1}) and (\ref{b2}) we get for 
the interaction density the expression: 
\be 
b_{1x}^{+}b_{2x}\simeq 
e^{i\sqrt{\pi}\sqrt{\xi}(-{\tilde  \phi}_1(x)+{\tilde  \phi}_2(x)+\phi_1(x)+\phi_2(x))}
=e^{i2\sqrt{\pi}\sqrt{\xi}\phi(x)},  
\label{b1b2}
\ee 
where we have introduced two new fields $\phi(x)$ and $\chi(x)$ defined 
according to the equations
\[
\phi(x)=\sqrt{\pi}(\Nh_{2}^{(1)}+\Nh_{1}^{(2)})(x), ~~~
\chi(x)=\sqrt{\pi}(\Nh_{1}^{(1)}+\Nh_{2}^{(2)})(x), 
\] 
or in terms of the dual fields 
\[
\phi(x)+\chi(x)=\phi_1(x)+\phi_2(x), ~~~~
\phi(x)-\chi(x)=-{\tilde \phi}_1(x)+{\tilde \phi}_2(x). 
\]
In terms of this new fields we get exactly the Lagrangian of the SG- model: 
\be
L=\frac{1}{2}(\partial_{\mu}\phi)^2+\frac{1}{2}(\partial_{\mu}\chi)^2+
C\mu^{\xi}h\cos(\beta\phi) 
\label{SG}
\ee
with the correct value of the constant $\beta=2\sqrt{\pi}\sqrt{\xi}$ 
or $\beta^2=8(\pi-\eta)$. In the equation (\ref{SG}) $\mu$ is the 
normalization point and the dimensionless constant $C$ 
before the term $\cos(\beta\phi)$ was found in \cite{Z} but 
cannot be fixed exactly in the framework of our approach because of 
the contribution of the operators $e^{\pm i2\eta N^{(1,2)}(x)}$ to the 
interaction term. The dimension of $\mu$ is equal to unity, while 
the dimension of the coupling constant $h$ is $2-\xi=2\eta/\pi$ so that 
the dimension of the factor $\mu^{\xi}h$ in eq.(\ref{SG}) is exactly 
equal to the $(mass)^2$ (see the expression for the physical 
mass $M$ which by definition has the dimension of mass in Section 4). 
Note that the constants in front of the operators (\ref{b1}),(\ref{b2}) 
depend on the normalization point $\mu$ in such a way that the 
lattice correlator does not depends on $\mu$. 
The auxiliary field $\chi(x)$ decouples from the 
SG- model. The direct evaluation of the constant $\beta$ is the 
main result of the present Letter. 
One can see from eq.(\ref{b1b2}) that as it should be, the effective 
low-energy theory at $M\neq L/2$ depends on the total number of Bosons $M$ 
in such a way that it leads to a twist boundary conditions for the field 
$\phi(x)$ of order $\Delta M=(M-L/2)$. In fact one should shift both 
the dual fields ${\tilde\phi}_{1,2}(x)$ (as can be seen from the Luttinger 
liquid relation for a system with twist) and the fields $\phi_{1,2}(x)$ 
in such a way that the boundary conditions for the Sine-Gordon model become 
$\beta(\phi(L)-\phi(0))=2\pi\Delta M$ which corresponds exactly to the 
boundary conditions for $\Delta M$ solitons.

\vspace{0.2in}

{\bf 4. Critical behaviour.}

\vspace{0.2in}

Let us calculate the physical mass of the soliton (dressed particle 
or hole) and the vacuum energy and compare the behaviour of this 
quantities as a functions of the coupling constant $h=e^{-\theta}$ 
with the general predictions of the perturbation theory in $h$. 
The Bethe Ansatz equations for the parameters 
$t_{\a}$, $\a=1,\ldots M$,  
which determine the common eigenstates of the transfer matrix (\ref{Z}) 
and the Hamiltonian (\ref{H}) have the standard form: 
\be
\left(\frac{\sh(t_{\a}-i\eta/2)}{\sh(t_{\a}+i\eta/2)}\right)^{L/2} 
\left(\frac{\sh(t_{\a}-\theta-i\eta/2)}{\sh(t_{\a}-\theta+i\eta/2)}
\right)^{L/2}=\prod_{\g\neq\a}
\frac{\sh(t_{\a}-t_{\g}-i\eta)}{\sh(t_{\a}-t_{\g}+i\eta)} 
\label{BA}
\ee
The solution of the equations (\ref{BA}) is similar to the solution 
of the corresponding equations for the XXZ- spin chain. 
In terms of the parameters $t_1,\ldots t_M$ the energy and the momentum 
of the eigenstates of the operator (\ref{Ham}) are 
\be
E=(\sin(\eta)/2)\sum_{\a}(\phi^{\prime}(t_{\a})+\phi^{\prime}(t_{\a}-\theta)), 
~~~P=\sum_{\a}(\phi(t_{\a})+\phi(t_{\a}-\theta)), 
\label{EP}
\ee
where the function $\phi(t)=(1/i)\ln(-\sh(t-i\eta/2)/\sh(t+i\eta/2))$. 
For the ground state the roots $t_{\a}$ are real and the corresponding 
density of roots $R(t)$ equals 
\be 
R(t)=\frac{1}{2}(R_0(t)+R_0(t-\theta)), ~~~~
R_0(t)=\frac{1}{2\eta\ch(\pi t/\eta)}. 
\label{R}
\ee
The calculation of the energy and the momentum of the single hole is 
quite standard and the result is analogous to that for the XXZ- spin chain: 
\be 
\epsilon(t)=(\sin(\eta)/2)2\pi(R_0(t)+R_0(t-\theta)), ~~~~
p^{\prime}(t)=2\pi(R_0(t)+R_0(t-\theta)), 
\label{ep}
\ee
where $t$ is the rapidity of the hole and the prime means the derivative 
over $t$. From the equation (\ref{ep}) in the limit 
$\theta\rightarrow\infty$ one can easily obtain the relativistic dispersion 
relation for the soliton: 
\be
\epsilon(t)=M\ch(\pi t/\eta), ~~~p(t)=M\sh(\pi t/\eta), ~~~
M=4\sqrt{v}e^{-\pi\theta/2\eta}.  
\label{disp}
\ee
The physical mass equals $M=4\sqrt{v}e^{-\pi\theta/2\eta}$ where  
$v=(\sin(\eta)/\eta)(\pi/2)$ - is the sound velocity of the single XXZ- 
spin chain. It appears in eq.(\ref{disp}) because as was shown in the 
previous section, the relativistic form of the Lagrangian (\ref{SG}) 
was obtained only after the corresponding rescaling of the space coordinate 
and time. Thus the physical mass $M$ is calculated. 
Note that once the physical mass is evaluated it also fix the canonical 
dimensions of all dimensional parameters in the Lagrangian written 
down in the continuous space and time (see Section 3). 
Now the energy of the ground state $E_0(h)$ for the Hamiltonian (\ref{Ham}) 
can be easily calculated. It is given by the following Fourier integral: 
\be
E_0(h)=\frac{1}{2}\left(\frac{\sin(\eta)}{2}\right)\int d\w e^{i\w\theta}
\frac{\sh(\w(\pi-\eta)/2)}{\sh(\w\pi/2)\ch(\w\eta/2)}. 
\label{int}
\ee 
The vacuum energy (\ref{int}) is divergent which means that $E_0(h)$ 
contains the terms of order $\sim h^2$ which is parametrically much larger 
than the constant term $\sim e^{-\pi\theta/\eta}$, and which should be 
subtracted to express (\ref{int}) in terms of the physical mass $M$. 
This is equivalent to the evaluation of the contribution of the pole of 
the factor $1/\ch(\w\eta/2)$ in the integrand. 
Thus in terms of the soliton mass $M$ the expression for the energy takes 
the form (one should take into account that our SG- model is defined at 
the interval $(0,L/2)$)
\be
E_0(h)=\frac{1}{4}M^{2}\ctg\left(\frac{\pi^2}{2\eta}\right) 
\label{E0}
\ee
in agreement with \cite{DDV},\cite{LZ}. 

Now we confirm our expression for $\beta$ and the critical behaviour 
of the mass gap and the ground state energy found from the exact solution. 
Consider the system of free massless Bose field $(H_0)$ perturbed by the 
relevant operator $V=h\sum_{x}V_x$. The perturbation theory in the 
coupling constant $h$ has the infrared divergences. To take them into 
account one has to sum up the whole perturbation theory series. 
In general for the ground and excited states we have the expression 
of the type 
\[
E(h)=V\frac{1}{E_0-H_0}V\left(1+U+U^2+\ldots\right), 
~~~U=\frac{1}{E_0-H_0}V\frac{1}{E_0-H_0}V. 
\]
To estimate the first term we write 
\[
\la V\frac{1}{E_0-H_0}V\ra\sim h^{2}\frac{1}{(1/L)}\sum_{i,j}
\la V_{i}V_{j}\ra\simeq L^{2}h^{2}\sum_x\frac{1}{x^d}\sim 
h^{2}L^{3-2d},  
\]
where $d$- is the scaling dimension of the operator $V_x$. 
Analogously we obtain $U\sim h^{2}L^{4-2d}$. We see that the operator 
$V$ is relevant provided $d<2$. Thus the ground state energy has the 
form $E_0(h)=h^{2}L^{3-2d}f(h^{2}L^{4-2d})$ with some unknown function 
$f(y)$. From the condition $E_0\sim L$ one can find the behaviour 
of $f(y)$ at large $y$. Thus we obtain the result: 
\be 
E_0(h)\sim h^{\frac{2}{2-d}}. 
\label{E}
\ee
Analogously for the mass gap we find: 
\be 
M\sim h^{\frac{1}{2-d}}. 
\label{M}
\ee
The equations (\ref{E}), (\ref{M}) are known also from the 
conformal line of arguments \cite{G}. In our case the scaling dimension 
$d=\beta^2/4\pi=\xi$ and we see that the equations (\ref{E}), (\ref{M}) 
are in agreement with the predictions (\ref{disp}), (\ref{E0}) 
(in our case $h\sim e^{-\theta}$ and $1/(2-d)=1/(2-\xi)=\pi/2\eta$). 
Thus the complete agreement of the perturbation theory estimates 
with the exact results (\ref{disp}),(\ref{E0}) is established. 
Note that using the BA results 
(\ref{disp}),(\ref{E0}) one can predict the correct value of the 
constant $\beta$ using the renormalization group arguments \cite{DS}. 
In our case it is interesting that the renormalization group 
arguments correctly fix the canonical dimension of the interaction 
term in the Lagrangian (\ref{SG}) (the anomalous dimension of the 
operator $\cos(\beta\phi)$ equals $\beta^2/4\pi=\xi$).

\vspace{0.2in}

{\bf 5. Conclusion.}

\vspace{0.2in}

In the present paper we have shown that the six-vertex model 
with alternating inhomogeneity parameters can be used to construct 
the local lattice integrable regularization of the Sine-Gordon 
model. We have shown {\it directly} that the system is equivalent 
to the two weakly coupled XXZ- spin chains and up to the 
irrelevant operators the interaction gives exactly the Sine-Gordon 
Lagrangian with the correct value of the parameter $\beta$. 
We compare the soliton mass and the vacuum energy obtained in the 
framework of the Bethe Ansatz with the general predictions of the 
perturbation theory for their power-law behaviour in the coupling 
constant. The direct Bosonization proposed in the present paper 
can be useful for study of the other integrable models of 
relativistic quantum field theory.


\begin{thebibliography}{99}

\bibitem{DV}
C.Destri, H.J.de Vega, Nucl.Phys.B 290 (1987) 363. 

\bibitem{RS}
N.Yu.Reshetikhin, H.Saleur, Nucl.Phys.B 419 (1994) 507. 

\bibitem{DS}
C.Destri, T.Segalini, Nucl.Phys.B 455 (1995) 759. 

\bibitem{LP}
A.Luther, I.Peschel, Phys.Rev.B9 (1974) 2911. 

\bibitem{H}
F.D.M.Haldane, Phys.Rev.Lett.47 (1981) 1840; J.Phys.C 14 (1981) 2585. 

\bibitem{Z}
Al.B.Zamolodchikov, Int.J.Mod.Phys.A 10 (1995) 1125. 

\bibitem{DDV} 
C.Destri, H.J.de Vega, Nucl.Phys.B 358 (1991) 251.

\bibitem{LZ} 
S.Lukyanov, A.Zamolodchikov, Nucl.Phys.B 493 (1997) 571. 

\bibitem{G}
A.O.Gogolin, A.A.Nersesyan, A.M.Tsvelik, ``Bosonizations and Strongly 
Correlated Systems'', Cambridge University Press, Cambridge, 1998. 



\end{thebibliography}
\end{document}